\documentstyle[epsfig,amssymb]{aipproc}

\begin{document}

\title{Creation and Decay of $\eta$-Mesic Nuclei}

\author{G.A.\ Sokol%
\thanks{Talk given at CIPANP-2000 (May 22-28, 2000, Quebec).
   E-mail: gsokol@x4u.lebedev.ru},
T.A.\ Aibergenov, A.V.\ Koltsov, A.V.\ Kravtsov, \\
Yu.I.\ Krutov, A.I.\ L'vov, L.N.\ Pavlyuchenko, \\
V.P.\ Pavlyuchenko, S.S.\ Sidorin}

\address{P.N. Lebedev Physical Institute,
  Leninsky Prospect 53, Moscow 117924, Russia}

\maketitle

\begin{abstract}

First experimental results on photoproduction of $\eta$-mesic nuclei are
analyzed.  In an experiment performed at the 1 GeV electron synchrotron of
the Lebedev Physical Institute, correlated $\pi^+n$ pairs arising from the
reaction
$$
  \gamma + {}^{12}{\rm C} \to N + {}_\eta(A-1)
      \to N + \pi^+ + n + (A-2)
$$
and flying transversely to the photon beam have been observed.  When the 
photon energy exceeds the $\eta$-meson production threshold, a distribution 
of the $\pi^+n$ pairs over their total energy is found to have a peak in 
the subthreshold region of the internal-conversion process $\eta p \to 
\pi^+ n$ which signals about formation of $\eta$-mesic nuclei.

\end{abstract}

The idea that a bound state of the $\eta$-meson and a nucleus (the
so-called $\eta$-mesic nucleus) can exist in Nature was put forward long
ago by Peng \cite{pen85} who relied on the first estimates of the $\eta N$
scattering length $a_{\eta N}$ obtained by Bhalerao and Liu \cite{bha85}.
Owing to ${\rm Re\,} a_{\eta N} > 0$, an average attractive potential
exists between slow $\eta$ and nucleons. This can result in binding $\eta
A$ systems, provided the life time of $\eta$ in nuclei is long enough
\cite{liu86}. Modern calculations \cite{bat98,gre97} predict a rather
strong $\eta N$ attraction which is sufficient for binding $\eta$ in all
nuclei with $A \ge 4$.

The very first experiments on searching for the $\eta$-mesic nuclei
performed at BNL \cite{chr88} and LAMPF \cite{lie88} gave negative results.
Meantime, studies of the reactions $p(d,{}^3{\rm He})\eta$
\cite{ber88,may96}, ${}^{18}{\rm O}(\pi^+,\pi^-){}^{18}{\rm Ne}$
\cite{joh93}, and $\vec d(d,{}^4{\rm He})\eta$ \cite{wil97} suggest that a
quasi-bound $\eta A$ state is formed in these reactions \cite{wil93,kon94}.

\begin{figure}[ht]
\centerline{\epsfig{file=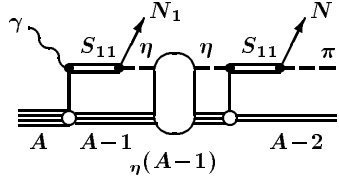,width=0.3\textwidth}}
\caption{\sl Mechanism of creation and decay of $\eta$-mesic nuclei.}
\label{fig:mechanism}
\end{figure}

In the present work we report on first results concerning formation of
$\eta$-mesic nuclei in photoreactions.  A very efficient trigger for
searching for $\eta$-mesic nuclei \cite{sok91} consists in detecting decay
products of the $\eta$-mesic nucleus, viz.\ a $\pi N$ pair produced in the
reaction $\eta N \to \pi N$ inside the nucleus.  Here $\eta$ itself is
produced at an earlier stage, in the reaction $\gamma N \to \eta N$ in our
case.  Both these reactions are mediated by the $S_{11}(1535)$ resonance
which affects also a propagation of the intermediate $\eta$ in the medium
(via multiple $\eta N$ rescattering) and leads to capturing slow $\eta$
into a bound state (Fig.~\ref{fig:mechanism}).  Formation of the bound
state of the $\eta$ and the nucleus becomes possible when the momentum of
the produced $\eta$ is small (typically less than 150 MeV/c).  This
requirement suggests photon energies $E_\gamma = 650{-}850$ MeV as most
suitable for creating $\eta$-nuclei.

$\pi N$ pairs emerging from $\eta$-mesic nucleus decays have an opening
angle $\langle \theta_{\pi N} \rangle = 180^\circ$ and specific kinetic
energies of their components (though smeared by the Fermi motion), $\langle
E_\pi \rangle \simeq 300$ MeV, $\langle E_n\rangle \simeq 100$ MeV. Among
four possible isotopic combinations $\pi^+ n$, $\pi^- p$, $\pi^0 n$, $\pi^0
p$ the first one is quite suitable for measuring energies of the particles.

\begin{figure}[ht]
\epsfig{file=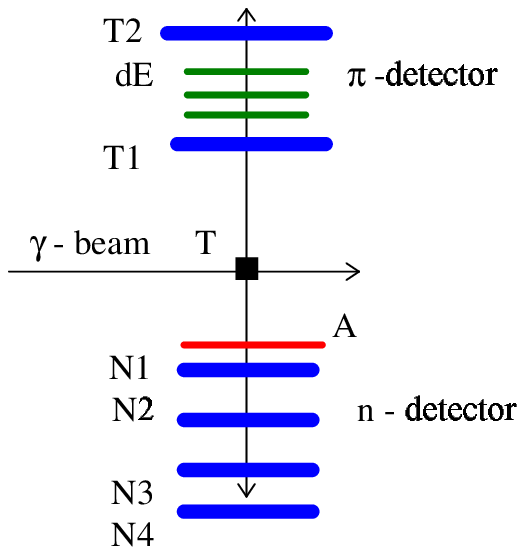,  width=0.28\textwidth, clip=}
\epsfig{file=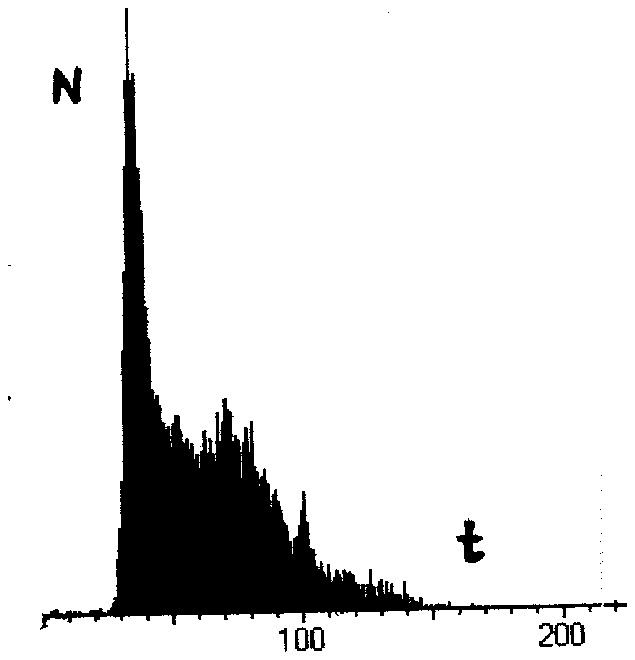, width=0.30\textwidth, clip=}
\epsfig{file=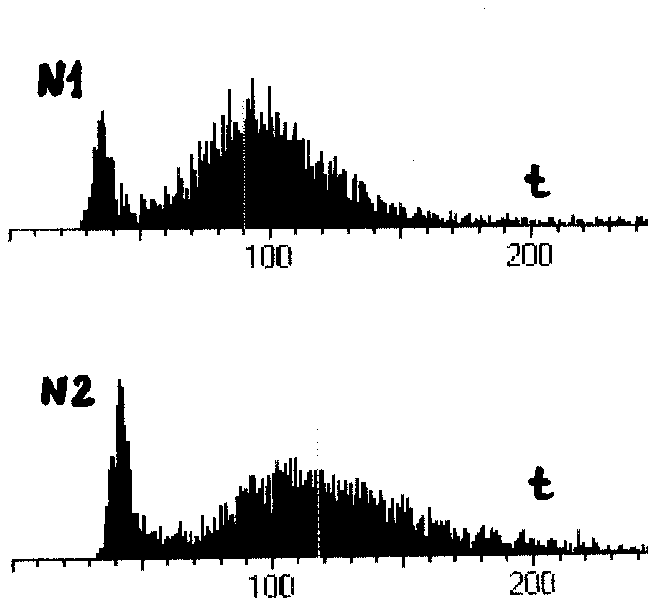, width=0.30\textwidth, clip=}
\bigskip
\caption{Layout of the experimental setup. Shown also
  time-of-flight spectra in the $\pi$ (left) and $n$ (right) spectrometers.}
\label{fig:setup}
\end{figure}

Accordingly, in an experiment performed at the 1 GeV electron synchrotron
of the Lebedev Physical Institute, correlated $\pi^+n$ pairs arising from
the reaction
\begin{equation}
  \gamma + {}^{12}{\rm C} \to N + {}_\eta(A-1)
      \to N + \pi^+ + n + (A-2)
\end{equation}
have been searched for.  An experimental setup (Fig.~\ref{fig:setup})
consisted of a carbon target $\rm \varnothing\, 4~cm \times 4~cm$ and two
time-of-flight scintillator spectrometers having a time resolution of
$\delta\tau \simeq 0.1$ ns. A plastic anticounter A of charged particles
(of the 90\% efficiency), placed in front of the neutron detectors,
and $dE/dx$ layers, placed between start and stop detectors in the pion
spectrometer, were used for a better identification of particles.

Strategy of measurements was as follows. Two bremsstrahlung-beam energies
were used, $E_{\gamma \rm max} = 650$ MeV and 850 MeV, i.e., well below and
well above $\eta$ production threshold on the free nucleon which is 707
MeV.  The first, ``calibration" run was performed at 650 MeV with the
spectrometers positioned at angles $\theta_n = \theta_\pi = 50^\circ$
around the beam. In that run, this was a quasi-free photoproduction $\gamma
p \to \pi^+ n$ which dominated the observed yield of the $\pi^+ n$
pairs.  Then, at the same ``low" energy 650 MeV, the spectrometers were
positioned at $\theta_n=\theta_\pi = 90^\circ$ (the ``background" run).
In such a kinematics, the quasi-free production did not contribute and the
observed count was presumably dominated by double-pion production.  At
last, the third run (the ``effect$+$background" run) was performed at the
same $90^\circ/90^\circ$ position, however with the higher photon beam
energy of 850 MeV, at which $\eta$ mesons are produced too.

\begin{figure}[htb]
\centerline{\epsfig{file=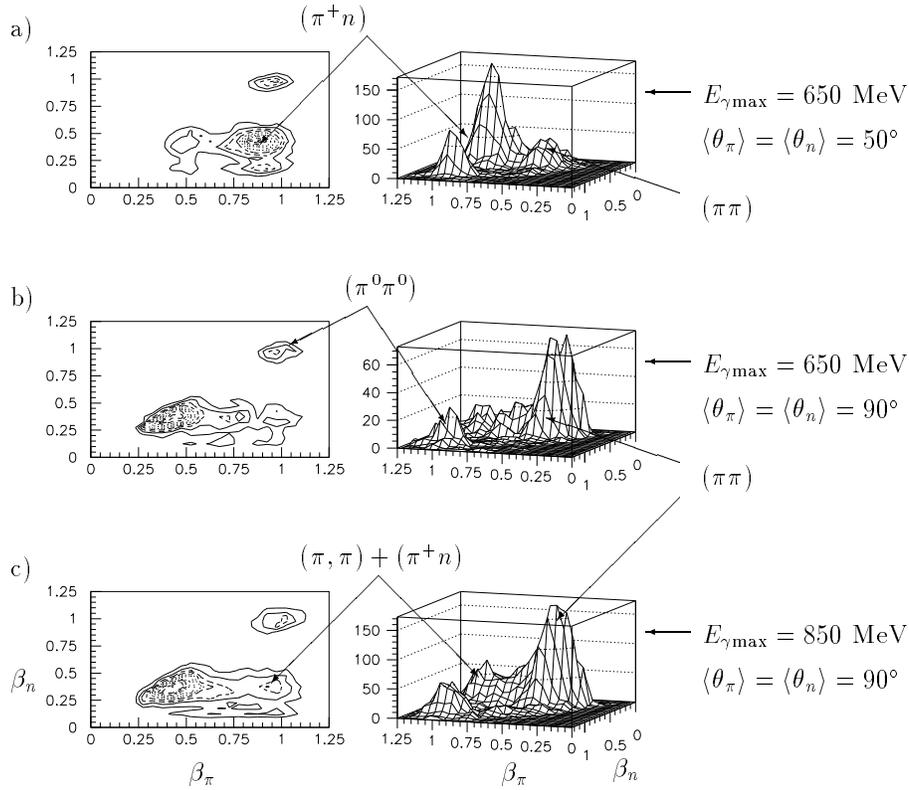,width=12cm}}
\bigskip
\caption{Raw $\pi^+ n$ event distributions over the particle velocities
     $\beta_n$ and $\beta_\pi$ for the ``calibration" run (a),
     ``background" run (b), and ``effect$+$background" run (c).}
\label{fig:betabeta-raw}
\end{figure}

In accordance with measured velocities of particles detected by the
spectrometers, all candidates to the $\pi^+ n$ events were separated into
three classes:  fast-fast (FF), fast-slow (FS), and slow-slow (SS) events.
The FF events mostly correspond to $\pi^0 \pi^0$ production which results
in hitting detectors by photons or $e^+/e^-$.  The FS events mostly emerge
from the $\pi^+ n$ pairs.  Comparing yields and time spectra in these runs
(and, in particular, using the SS events for extrapolating and subtracting
a background), we have found a clear excess of the FS events which appeared
when the photon energy exceeded $\eta$ production threshold. The total
cross section of photoproduction of such excess pairs, averaged over the
photon energy range $650{-}850$ MeV, was found to be about $\sigma_{\rm
tot}(\pi^+n) \simeq 10$ $\mu$b.  See Ref.\ \cite{sok98} for more details.
In the present work a further analysis of the excess FS events is done and
their energy characteristics are determined.

\begin{figure}[htb]
\centerline{\epsfig{file=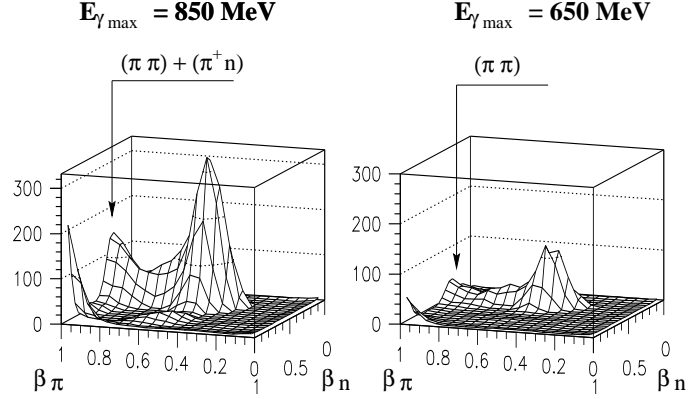,width=9cm}}
\caption{Corrected two-dimensional distributions over the velocities
    $\beta$ of the $\pi ^+ n$ events
    with the end-point energy of the bremsstrahlung spectrum
    $E_{\gamma \rm max} = 850$ and 650 MeV.}
\label{fig:betabeta-inverse}
\end{figure}

In order to find kinetic energies of the neutron and pion, the velocities
$\beta_i = L_i/ct_i$ of both the particles have to be determined.  They are
subject to fluctuations stemming from errors $\delta t_i$ and $\delta L_i$
in the time-of-flight $t_i$ and the flight base $L_i$.  Such fluctuations
are clearly seen in the case of the ultra-relativistic FF events which have
experimentally observed velocities close but not equal to 1 (see Fig.\
\ref{fig:betabeta-raw}).  Therefore, an experimental $\beta$-resolution of
the setup can be directly inferred from the FF events.  Then, using this
information and applying an inverse-problem statistical method described in
Ref.\ \cite{pav83}, one can unfold the experimental spectrum, obtain a
smooth velocity distribution in the physical region $\beta_i < 1$
(Fig.~\ref{fig:betabeta-inverse}), and eventually find a distribution of
the particle's kinetic energies $E_i = M_i [ (1-\beta_i^2)^{-1/2} - 1 ]$.
Finding $E_i$, we introduced corrections related with average energy losses
of particles in absorbers and in the detector matter.  It is worth to say
that the number of the $\pi^+ n$ FS events visibly increases when the
photon beam energy becomes sufficient for producing $\eta$ mesons.

\begin{figure}[ht]
\unitlength=1mm
\begin{picture}(100,55)(0,0)
\put(40,48){\small $E_{\gamma \rm max}=850$ MeV}
\put(90,48){\small $E_{\gamma \rm max}=650$ MeV}
\centerline{
\epsfxsize=4.5cm\epsfbox[157 322 443 613]{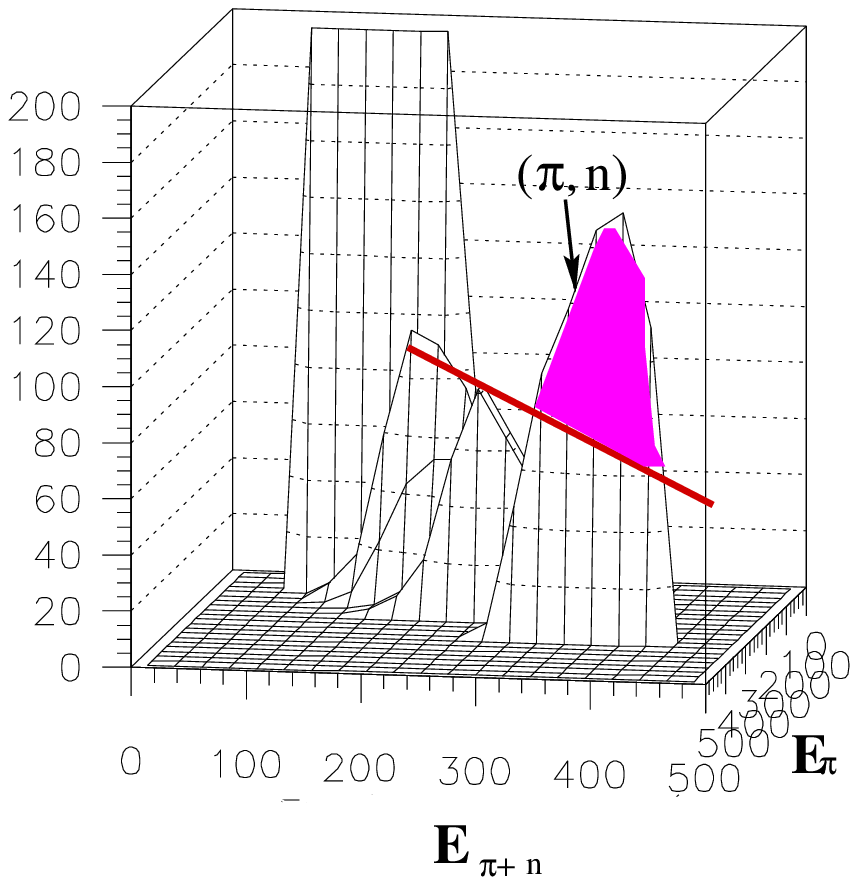}
\epsfxsize=4.5cm\epsfbox[16 325 304 670]{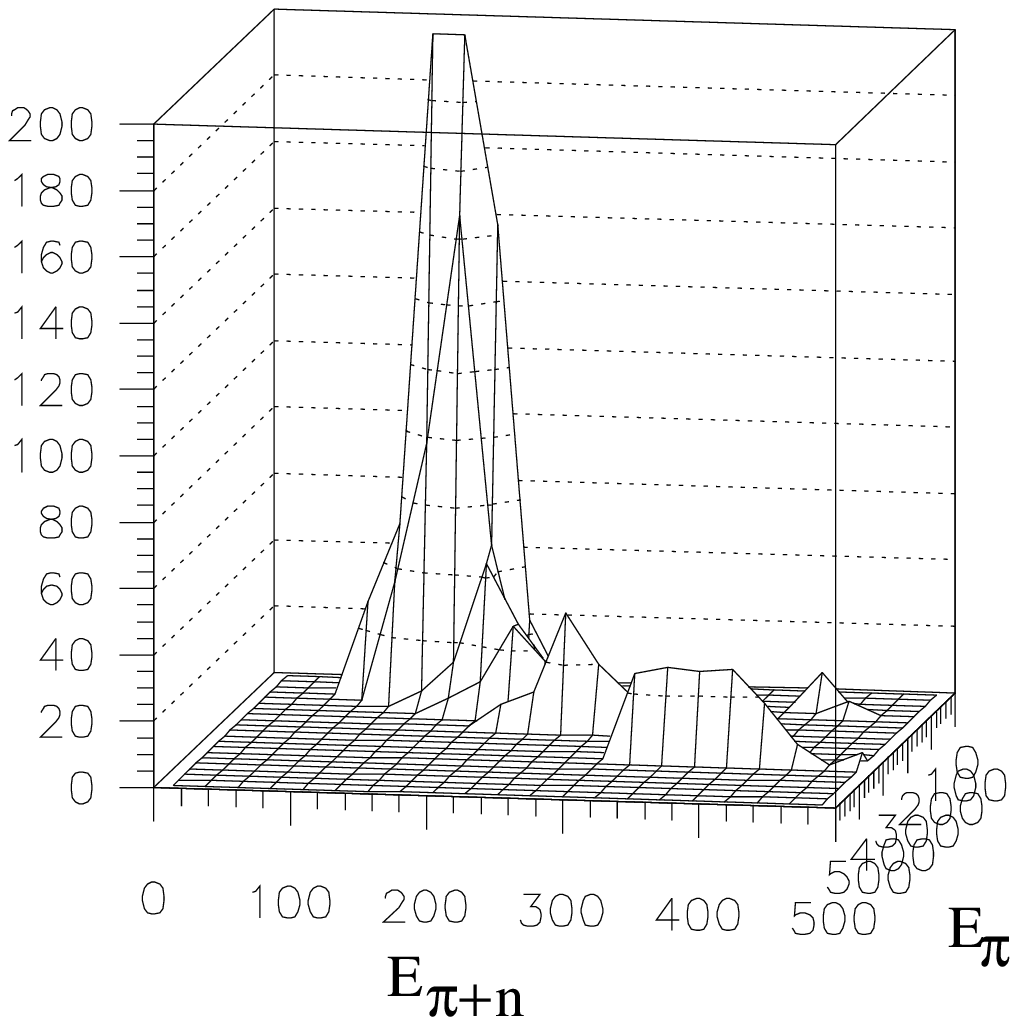}}
\end{picture}
\caption{Distribution over the total kinetic energy of the $\pi^+n$ pairs
      for the ``effect$+$background" run (the left panel) and for the
      ``background" run (the right panel) obtained after
      unfolding the raw spectra.}
\label{fig:Etot-2dim}
\end{figure}

\begin{figure}[htb]
\centerline{\epsfig{file=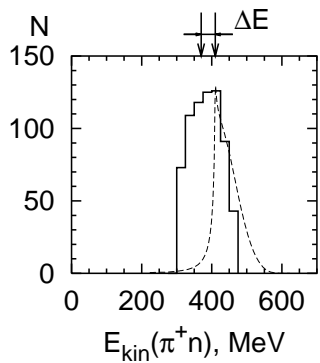,width=0.35\textwidth}}
\bigskip
\caption{Distribution over the total kinetic energy of the $\pi^+n$ pairs
     after subtraction of the background. Arrows indicate threshold
     in the reaction $\eta N \to \pi N$, i.e.\ 408 MeV, and the weighted
     center of the histogram. For a comparison, a product of free-particle
     cross sections of $\gamma N \to \eta N$ and $\eta N \to \pi N$
     \protect\cite{gre97} is shown with the dashed line (in arbitrary units).}

\label{fig:Etot-1dim}
\end{figure}

Of the most interest is the distribution of the $\pi^+ n$ events over their
total energy $E_{\rm tot} = E_n + E_\pi$, because creation and decay of
$\eta$-mesic nuclei is expected to produce a relatively narrow peak in
$E_{\rm tot}$ of the width $\sim 50{-}70$ MeV (see, e.g.,
\cite{sok98,lvo98}). Such a peak was indeed observed: see Fig.\
\ref{fig:Etot-2dim}, in which an excess of the FS events appears when the
photon energy exceeds the $\eta$-production threshold. Subtracting a smooth
background, we have found a 1-dimensional energy distribution of the $\pi^+
n$ events presumably coming from (bound) $\eta$ decaying in the nucleus,
see Fig.\ \ref{fig:Etot-1dim}.  The experimental width of this distribution
is about 100 MeV, including the apparatus resolution. Its center lies by
$\Delta E = 40$ MeV below the energy excess $m_\eta-m_\pi = 408$ MeV in the
reaction $\eta N \to \pi N$, and it is well below the position of the
$S_{11}(1535)$ resonance too.  Up to effects of binding of protons
annihilated in the decay subprocess $\eta p \to \pi^+ n$, the value $\Delta
E$ characterizes the binding energy of $\eta$ in the nucleus. The width of
that peak is determined both by the width of the $\eta$-bound state and by
the Fermi motion.

\begin{figure}[htb]
\unitlength=1mm
\begin{picture}(100,40)(0,0)
\put(030,30){\small $N$}
\put(115,01){\small $p_\perp$, MeV}
\put(042,37){\small $E_{\gamma \rm max}=850$ MeV}
\put(080,37){\small $E_{\gamma \rm max}=650$ MeV}
\centerline{\epsfig{file=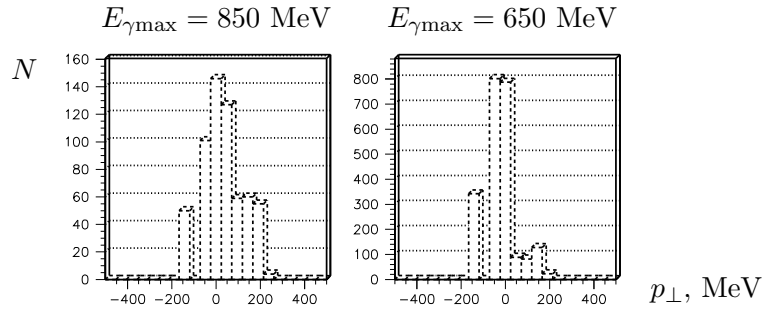,width=0.5\textwidth}}
\end{picture}
\bigskip
\caption{Distribution over the total transverse momentum $p_\perp$ of $\pi^+n$
    pairs for the ``effect$+$background" run (the left panel)
    and for the ``calibration" run (the right panel).}
\label{fig:momentum}
\end{figure}

Whereas the fixed opening angle $\theta_{\pi n}=180^\circ$ chosen in the
kinematics with $\theta_n=\theta_\pi = 90^\circ$ selects $\pi^+n$ pairs
carrying a low total momentum in the direction of the photon beam, an
independent check of the transverse momentum $p_\perp = p_\pi - p_n$ is
meaningful.  The corresponding distribution is shown in Fig.\
\ref{fig:momentum}.  On the top of a background, there is a narrower peak
in $p_\perp$ having a width compatible with the Fermi momentum of nucleons
in the nucleus.

In conclusion, an excess of correlated $\pi^+ n$ pairs with the opening
angle $\langle \theta_{\pi N} \rangle = 180^\circ$ has been experimentally
observed when the energy of photons exceeded the $\eta$-production
threshold. A distribution of the pairs over their total kinetic energy was
found to have a peak lying below threshold of the elementary process $\pi
N\to \eta N$.  A narrow peak is also found in the pair's distribution over
their total transverse momentum. All that suggests that these $\pi^+ n$
pairs arise from creation and decay of captured bound $\eta$ in the
nucleus, i.e., they arise through the stage of formation of an $\eta$-mesic
nucleus.

The work was supported by RFBR grant 99-02-18224.

\end{document}